\documentclass[runningheads]{llncs}
\usepackage{graphicx}
\usepackage{tabularx}
\usepackage{subcaption}
\usepackage{hyperref}
\hypersetup{colorlinks=true,citecolor=blue,linkcolor=blue}

\newcolumntype{L}[1]{>{\raggedright\arraybackslash}p{#1}}
\newcolumntype{C}[1]{>{\centering\arraybackslash}p{#1}}
\newcolumntype{R}[1]{>{\raggedleft\arraybackslash}p{#1}}

\begin{document}
\title{Deep Learning for Pneumothorax Detection and Localization in Chest Radiographs}
\titlerunning{Pneumothorax Detection and Localization in Chest Radiographs}

\author{Andr{\'e} Goo{\ss}en\inst{1}\and Hrishikesh Deshpande\inst{1}\and Tim Harder\inst{1}\and Evan Schwab\inst{2}\and Ivo Baltruschat\inst{3}\and Thusitha Mabotuwana\inst{4}\and Nathan Cross\inst{5}\and Axel Saalbach\inst{1}}
\authorrunning{A. Goo{\ss}en et al.}

\institute{Digital Imaging, Philips Research, Hamburg, Germany\and Clinical Informatics Solutions and Services, Philips Research, Cambridge, USA\and Institute for Biomedical Imaging, Hamburg University of Technology, Germany\and Radiology Solutions, Philips Healthcare, Bothell, USA\and Department of Radiology, University of Washington Medical Center, Seattle, USA}

\maketitle


\begin{abstract}
Pneumothorax is a critical condition that requires timely communication and immediate action. In order to prevent significant morbidity or patient death, early detection is crucial. For the task of pneumothorax detection, we study the characteristics of three different deep learning techniques: (i) convolutional neural networks, (ii) multiple-instance learning, and (iii) fully convolutional networks. We perform a five-fold cross-validation on a dataset consisting of 1003 chest X-ray images. ROC analysis yields AUCs of 0.96, 0.93, and 0.92 for the three methods, respectively. We review the classification and localization performance of these approaches as well as an ensemble of the three aforementioned techniques.
\end{abstract}

\section{Introduction}
In many institutions, the ability to prioritize specific imaging exams is made possible by use of stat or emergent labeling. However, because of overuse and misuse of these labels, a radiologist often has difficulties prioritizing exams with more medically significant findings. As a result, an automated system to triage positive critical findings should improve the management of patients.
Such a functionality could not only help in bringing attention to the critically ill patient, but also help the radiologist to better manage his time reading the exams.
Timely communication of critical findings, in this manner, is endorsed by the American College of Radiology (ACR) as they have defined three categories of findings: \emph{Category~1}: Communication within minutes, \emph{Category~2}: Communication within hours, \emph{Category~3}: Communication within days. Immediate actions have to be taken, especially for the \emph{Category~1} findings, in order to prevent significant morbidity or patient death. \emph{Category~1} findings include - amongst others - pneumothorax \cite{larson2014actionable}.

Pneumothorax is a lung pathology that is associated with abnormal collection of air in the pleural space between the lung and the chest wall.
It can result from a variety of etiologies including chest trauma, pulmonary disease, and spontaneously. Pneumothorax can be life-threatening and is considered an emergency in intensive care, requiring prompt recognition and intervention \cite{ptx_icu}.

Deep learning is currently the method of choice for numerous tasks in computer vision such as image classification.
With the availability of large datasets and advanced compute resources, deep learning has achieved a performance on par with the medical professionals in tasks such as diabetic retinopathy detection \cite{dl_med1} and skin cancer classification \cite{dl_med2}.

In this paper, we investigate and evaluate three deep learning architectures for the detection and localization of pneumothorax in chest X-ray images.

\section{Methods}
\label{sec:Methods}

\paragraph{Convolutional Neural Networks (CNNs)}
\label{sec:Methods_CNN}
are the most commonly employed network architectures for image classification. They have been successfully used in a broad range of applications from computer vision to medical image processing \cite{he2016deep,wang2017chestx} and can be optimized in an end-to-end fashion.

Initial work in the medical domain focused predominantly on the re-use of deep learning networks from the computer vision domain (transfer learning). This is achieved either in terms of pre-trained networks, which are used as feature extractors, or by means of fine-tuning techniques, i.e.~the adaptation of an existing network to a new application or domain.
Promising results for X-ray image analysis have been obtained already by means of features derived from pre-trained networks \cite{lopes2017pre}.

\begin{figure}[tb]
\centering
\includegraphics{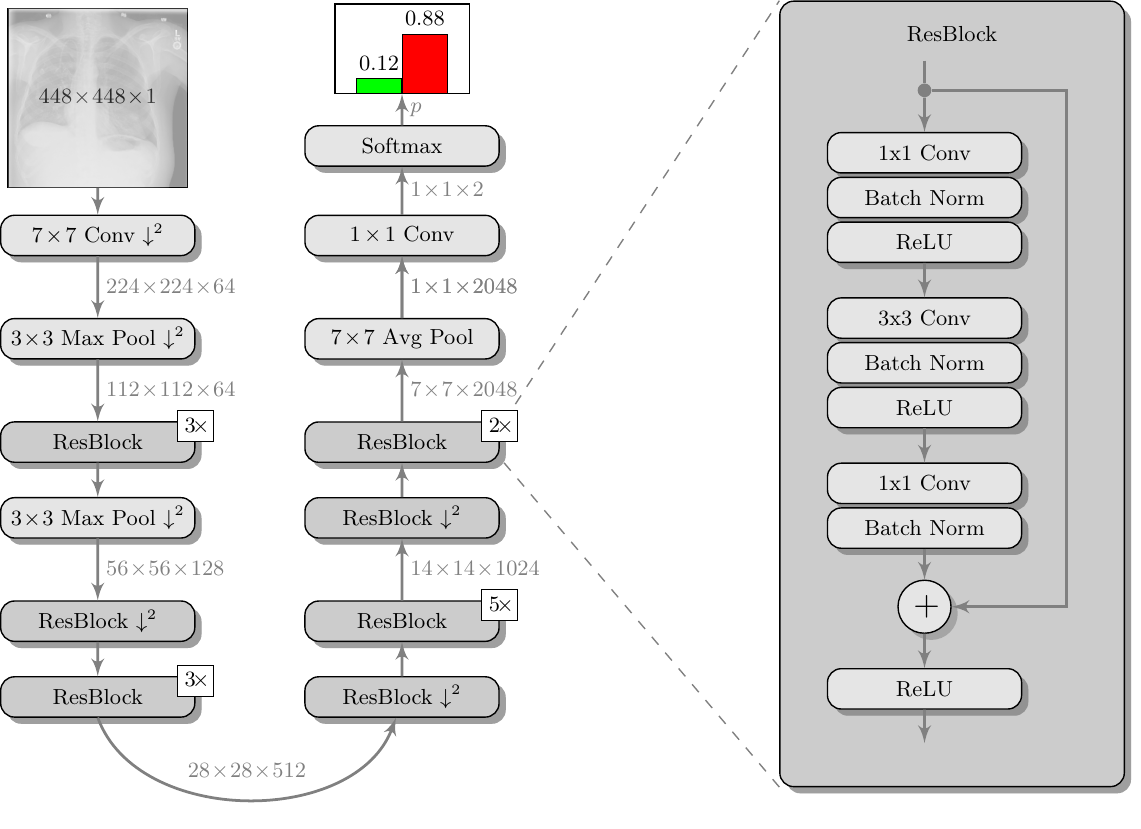}
\caption{ResNet-50 architecture of Baltruschat et~al.~\cite{baltruschat2018comparison} adapted for end-to-end binary pneumothorax classification. $\downarrow^2$ denotes a downsampling operation using a stride of 2. Repeating ResBlocks have been collapsed for readability.}
\label{fig:CNN}
\end{figure}

In the following method, a specific network architecture - a residual network - is employed. We use a variant of the ResNet-50 architecture \cite{he2016deep} with a single input channel and an enlarged input size of $448 \times 448$, which allows to leverage the higher spatial resolution of X-ray data, e.g.~for the detection of small structures \cite{baltruschat2018comparison}. Therefore, an additional pooling layer was introduced after the first bottleneck block (cf.~Fig.~\ref{fig:CNN}).
The network was trained on the NIH ChestX-ray14 dataset \cite{wang2017chestx} to predict 14 pathologies. For the task of pneumothorax detection, the dense layer for the prediction of pathologies was replaced by a new layer for binary classification.

\paragraph{Multiple-Instance Learning (MIL)}
\label{sec:Methods_MIL}
\cite{dietterich1997solving} provides a joint classification and localization, while only requiring the image-level labels for training. This approach may be advantageous in medical applications \cite{yan2016multi} where pixel-level labels are difficult to obtain and often require experts to perform the annotation.

To produce local predictions in the image, the full resolution chest X-ray images are partitioned into $N$ overlapping image patches, forming a bag. The goal is to produce a binary classification for each patch where a patch is defined as positive ($p_i=1$) if it contains pneumothorax and negative ($p_i=0$) if it does not contain pneumothorax.

\begin{figure}[tb]
\centering
\includegraphics{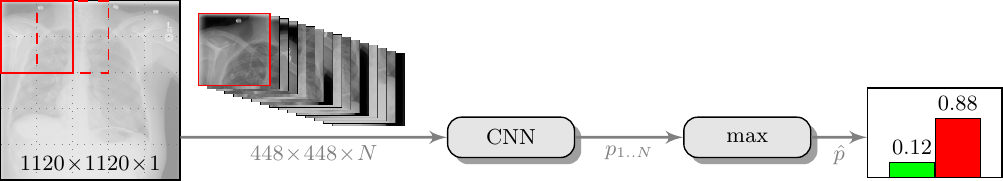}
\caption{The proposed Multiple-Instance Learning architecture, using the CNN as patch classifier, for joint pneumothorax classification and localization.}
\label{fig:MIL}
\end{figure}

Using the bag labels, it is known that all the patches in a non-pneumothorax image will necessarily be negative. On the other hand, at least one of the patches in a pneumothorax image must contain the pathology and therefore be a positive patch. MIL attempts to learn the fundamental characteristics of the local pathology by automatically differentiating between normal and abnormal characteristics of the chest X-ray.
Using these assumptions, MIL provides a mechanism to relate patch-level predictions, $p_{1..N}$, to bag labels by taking the maximum patch score $\hat{p}$ as the image-level classification.

Fig.~\ref{fig:MIL} shows a schematic of the proposed architecture. In this architecture, we use the previously discussed ResNet-50 network as patch classifier.

\begin{figure}[tb]
\centering
\includegraphics{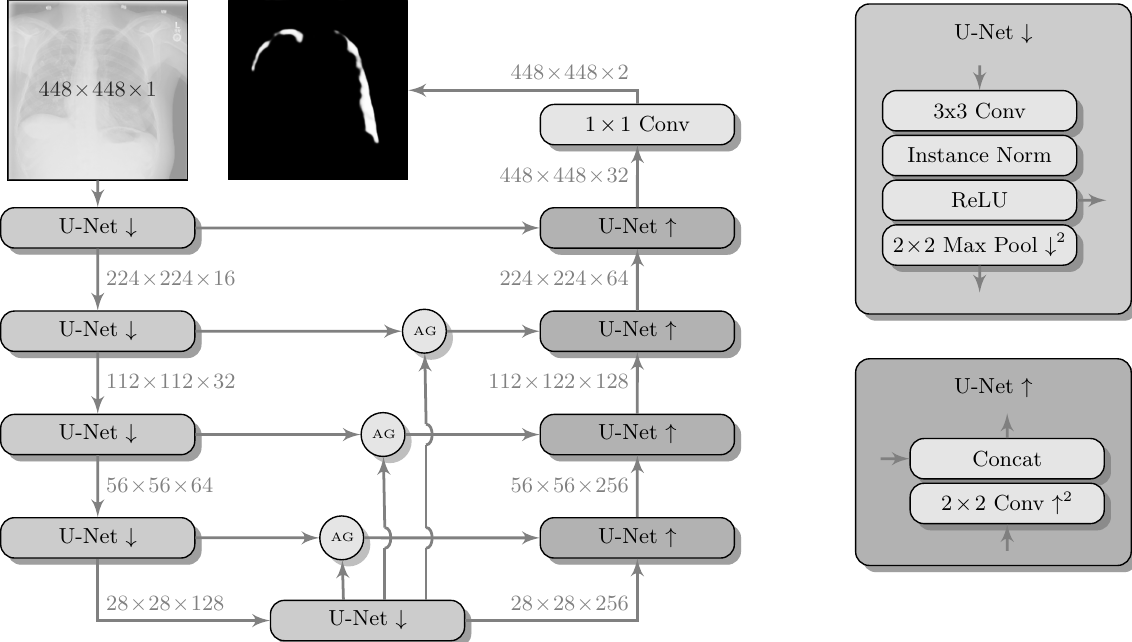}
\caption{The proposed FCN architecture using a four-layer U-Net \cite{ronneberger2015u} with Attention Gates (AG) in the skip connection \cite{oktay2018attention}.}
\label{fig:FCN}
\end{figure}

\paragraph{Fully Convolutional Networks (FCNs)}
\label{sec:Methods_FCN}
are more advanced network architectures, that have been developed for semantic segmentation, i.e.~pixel-level classification. The most commonly employed network in this context is the U-Net \cite{ronneberger2015u}, which consists of a contracting path resembling a CNN, for the integration of context information, and a corresponding expanding path. This allows to obtain  probability maps of the same size as the input image, facilitating the image localization.
For this experiment, we employ a U-Net with four layers per path and \emph{Attention Gates} \cite{oktay2018attention}. Attention gates have been proposed as an alternative to a detection component and they are employed in order to facilitate the segmentation of an object of interests. Furthermore, the proposed architecture uses instance normalization instead of commonly used batch normalization in order to harmonize the input data (cf. Fig.~\ref{fig:FCN}).

In contrast to CNNs, the FCN approach requires pixel-level annotations during the training and predicts probability values for each pixel during the application phase.
Therefore, it does not directly generate the image-level label, but requires an additional post-processing step.
In the scope of this study, we define the area of the detected pneumothorax as a classification measure. Although such measure is biased towards the detection of large pneumothorax regions, it is conceptually simple and favors the detection of reliable candidates.

\section{Experiments}
\label{sec:Experiments}
The data used in the following experiments consists of DICOM X-ray images, obtained from the University of Washington Medical Center and affiliated institutions, centered in Seattle by scanning radiology reports from the last three years. Inclusion criteria were: (i)~Digital Radiography (DR) images, (ii)~Chest radiographs, (iii)~Posterior-anterior or anterior-posterior view position, (iv)~Adult patients. Any personal health information was removed. Image-level labels were derived from natural-language processing based analysis of the reports. Cases were partially reviewed by a radiologist to confirm appropriate finding in the report's impression section and this represented a critical finding. The resulting dataset contained 1003 images: 437 with pneumothorax, 566 with a different or no abnormality detected. We generated pixel-level annotations of the pneumothorax region for 305 of the positive cases. For training and evaluation, we divided the dataset into five cross-validation splits of similar size, such that images of the same patient resided in the same split.

\begin{figure}[tb]
        \centering
        \includegraphics{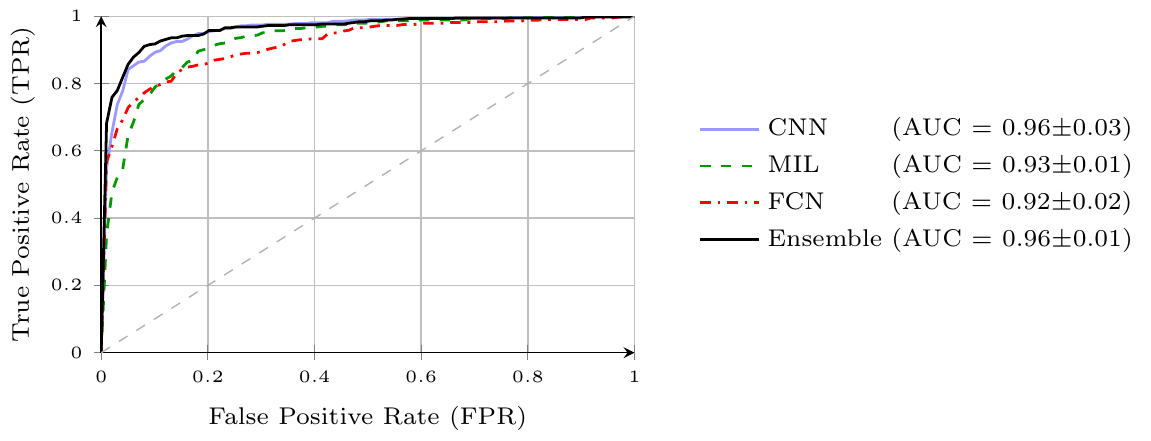}
        \caption{Averaged ROC curves over five splits for all methods and an ensemble.}
        \label{fig:ROC}
\end{figure}

To increase the variability of the available data, we augmented the dataset by translating, scaling, rotating, horizontal flipping, windowing, and adding Poisson noise. Input images for CNN and FCN have been created by cropping a centered patch of $448\times448$ from the original images resized to $480\times480$. For MIL we cropped overlapping patches out of the image resized to $1120\times1120$ (cf. Fig.~\ref{fig:MIL}). In training, we used the Adam optimizer with default parameters $\beta_1=0.9$ and $\beta_2=0.999$, a batch size of 16, and exponentially decreasing learning rate (LR). Refer to Table~\ref{table:experiments} for an overview of the parameters and to Fig.~\ref{fig:ROC} for the receiver operating characteristic (ROC) analysis we performed to assess the model performance.

\begin{table}[tb]
\caption{Experimental set-up for the training of the three networks. The four last rows indicate whether the network uses image-level or pixel-level labels for training and whether it provides classification or localization, respectively.}
\begin{center}
\begin{tabular}{| R{4cm} || C{2cm} | C{2cm} | C{2cm} |}
\hline
& \textbf{CNN} & \textbf{MIL} & \textbf{FCN} \\
\hline
\textbf{number of parameters\,} & 24M & 24M & 2.1M \\
\textbf{input size\,} & 448$\times$448 & 448$\times$448 & 448$\times$448 \\
\textbf{batch size\,} & 16 & 16 & 16 \\
\textbf{learning rate\,} & $10^{-4}$ & $10^{-5}$ & $10^{-4}$ \\
\textbf{epochs\,} & 40 & 30 & 400 \\
\hline
\textbf{image-level labels\,} & + & + & - \\
\textbf{pixel-level labels\,} & - & - & + \\
\hline
\textbf{classification\,} & + & + & $\circ$ \\
\textbf{localization\,} & - & $\circ$ & + \\
\hline
\end{tabular}
\label{table:experiments}
\end{center}
\end{table}

\paragraph{CNN:}
\label{sec:Experiments_CNN}
The pre-trained ResNet-50 was fine-tuned with an initial LR of $10^{-4}$ for 40 epochs. For testing, an average five crop response of the model, i.e.~center and all four corners, was used for the classification purpose. Very high and stable results can be reported, with area under curve (AUC) values of 0.96$\pm$0.03.

\begin{figure}[tb]   
        \centering
		\begin{subfigure}[b]{0.25\textwidth}
			\centering
			\includegraphics[width=0.95\linewidth]{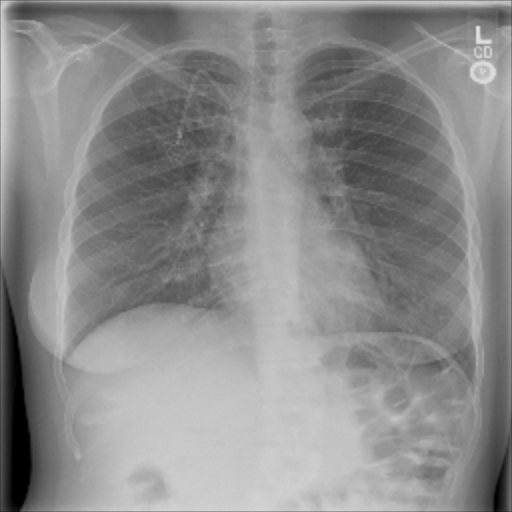}\vspace{0.2em}
			\includegraphics[width=0.95\linewidth]{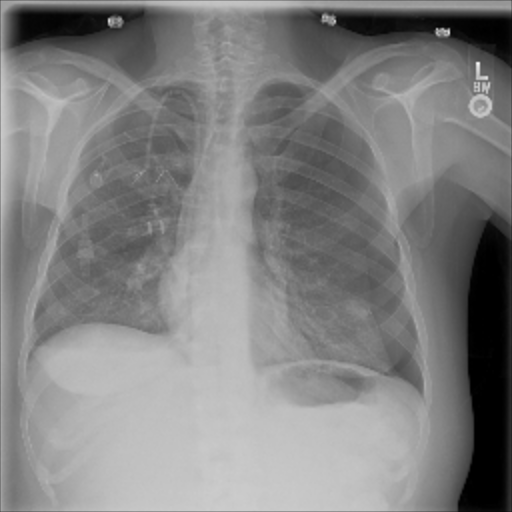}\vspace{0.2em}
			\includegraphics[width=0.95\linewidth]{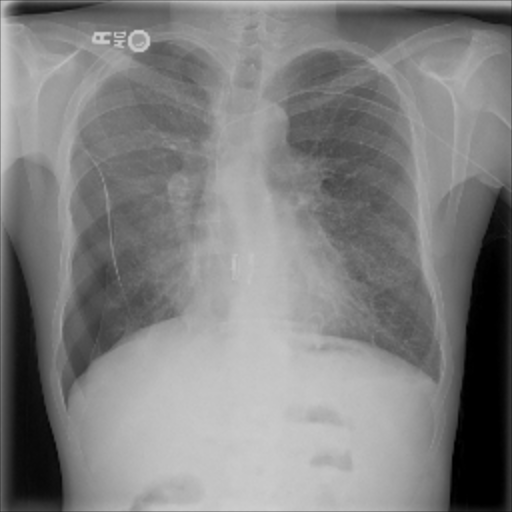}
			\caption{input image}
			\label{fig:localization_image}
		\end{subfigure}%
		\begin{subfigure}[b]{0.25\textwidth}
			\centering
			\includegraphics[width=0.95\linewidth]{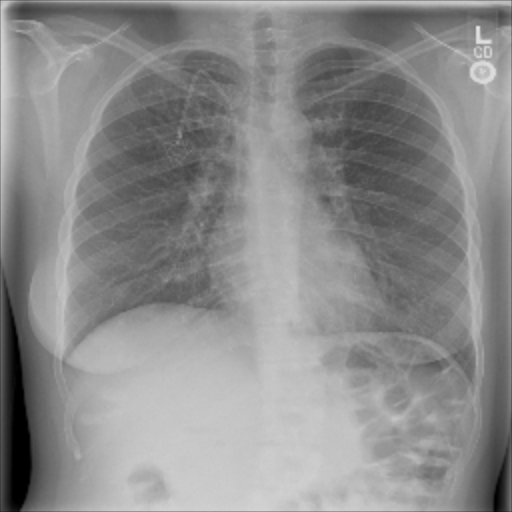}\vspace{0.2em}
			\includegraphics[width=0.95\linewidth]{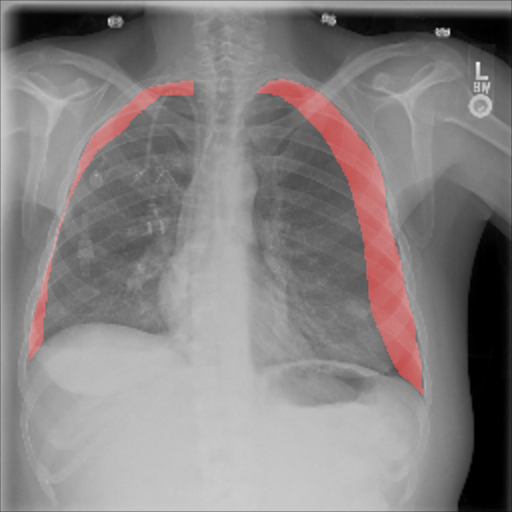}\vspace{0.2em}
			\includegraphics[width=0.95\linewidth]{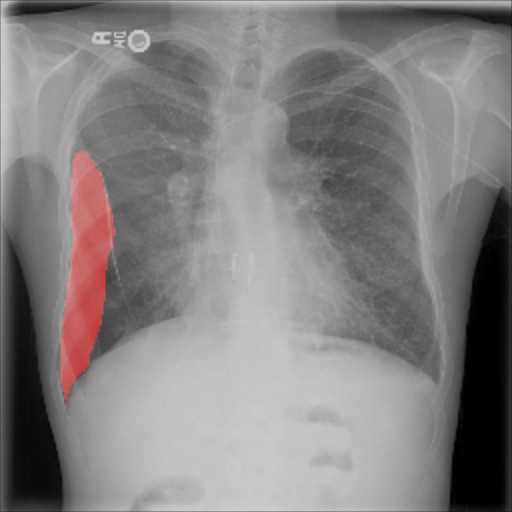}
			\caption{ground truth}
			\label{fig:localization_gt}
		\end{subfigure}%
		\begin{subfigure}[b]{0.25\textwidth}
			\centering
			\includegraphics[width=0.95\linewidth]{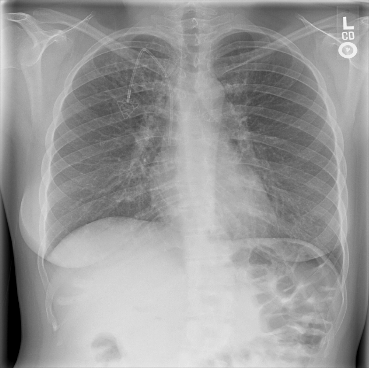}\vspace{0.2em}
			\includegraphics[width=0.95\linewidth]{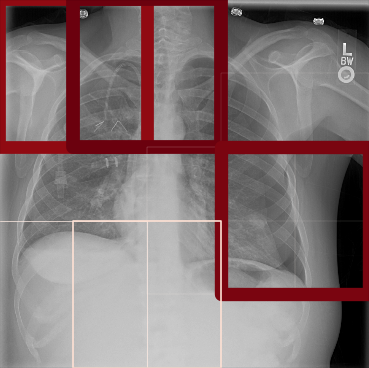}\vspace{0.2em}
			\includegraphics[width=0.95\linewidth]{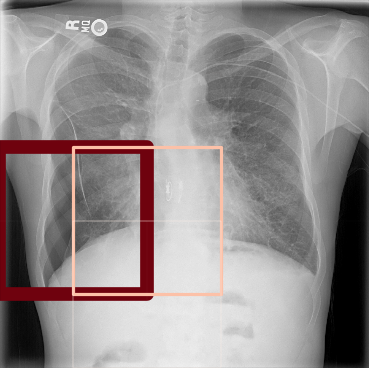}
			\caption{MIL}
			\label{fig:localization_MIL}
		\end{subfigure}%
		\begin{subfigure}[b]{0.25\textwidth}
			\centering
			\includegraphics[width=0.95\linewidth]{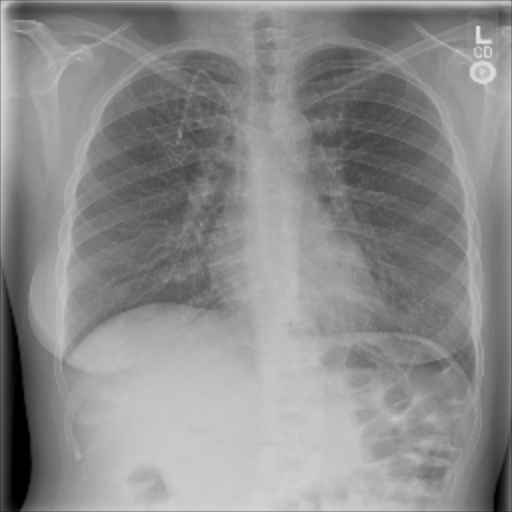}\vspace{0.2em}
			\includegraphics[width=0.95\linewidth]{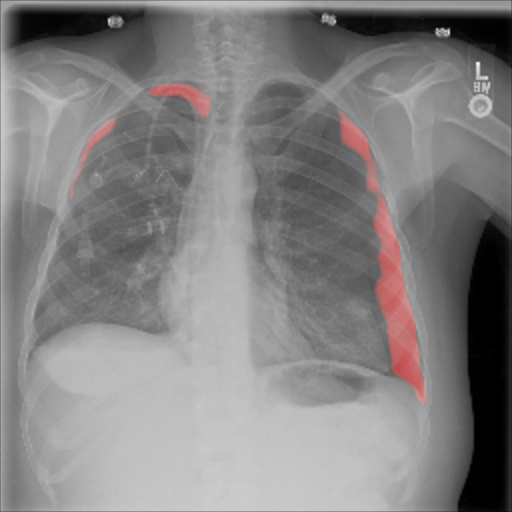}\vspace{0.2em}
			\includegraphics[width=0.95\linewidth]{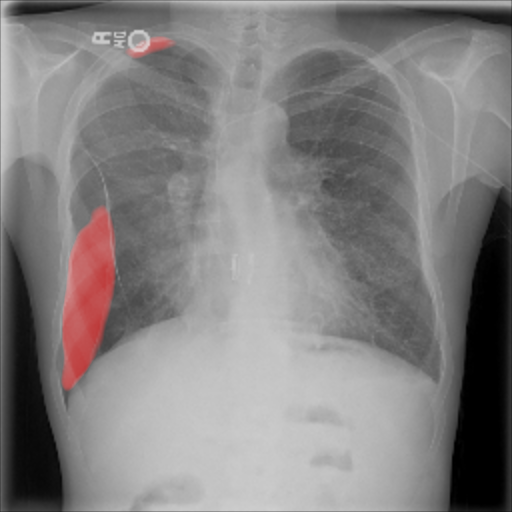}
			\caption{FCN}
			\label{fig:localization_FCN}
		\end{subfigure}		
	
    	\caption{Localization compared to manual annotation for a normal and two pneumothorax cases using Multiple-Instance Learning (MIL, thicker frames denote higher patch scores $p_i$) and a Fully Convolutional Network (FCN).}
   		\label{fig:examples}
\end{figure}

\paragraph{MIL:}
\label{sec:Experiments_MIL}
The pre-trained ResNet-50 was also employed as the patch-level classifier within the MIL approach. We chose the binary cross-entropy between the maximum patch score and the image-level label as the loss function. The batch size was selected as the number of $N=16$ patches per image. We trained with an initial LR of $10^{-5}$ for 30 epochs and achieved an average AUC of 0.93$\pm$0.01 using this method. High patch scores (indicated by thicker red frames, cf.~Fig.~\ref{fig:localization_MIL}) give a hint on the location of the pneumothorax.

\paragraph{FCN:}
\label{sec:Experiments_FCN}
As pixel-level ground truth annotations were available only for a subset of the images, 871 images in total were used for training the FCN for 400 epochs. As a loss function, a weighted cross entropy (25.0 for pneumothorax pixels and 0.5 for non-pneumothorax pixels in order to account for the smaller size of pneumothorax regions) was employed at pixel-level with an initial LR of $10^{-4}$.
With an average AUC of 0.92$\pm$0.02, the overall performance of this method is worse than the CNN and MIL. On the other hand, the FCN generates pixel-level probabilities (cf.~Fig.~\ref{fig:localization_FCN}), which indicate the location of the pneumothorax. The average Dice coefficient for positively classified cases is $54.2\%$.

\paragraph{Ensemble Learning:}
\label{sec:Experiments_Ensemble}
As can be seen from the previous sections, the different methods, that have been investigated, have their own advantages and disadvantages. However, looking at the performance, the errors made by different architectures do not necessarily coincide. Therefore, we investigated ensemble techniques, using linear combinations of the individual methods. The best parameter combination was identified using exhaustive search. The best ensemble of CNN, FCN, and MIL achieves the highest overall AUC of 0.965 (cf.~Fig.~\ref{fig:ROC}), but does not significantly (at $p<0.05$) outperform the CNN. CNN and FCN achieve best results amongst combining two techniques with an AUC of 0.962.

\section{Discussion}
Using the average AUC as a performance criterion, we achieved very stable results with values between 0.92 and 0.96 for all methods.
These results indicate a very good overall performance of the algorithms.

The AUC provides little information about the performance in different areas of the ROC space. Particularly for the worklist prioritization, it could be argued that an operating point with a low false positive rate (FPR) is of most relevance.
Even algorithms with a moderate true positive rate (TPR) could improve the clinical workflow compared to a sequential reading. In contrast, the reading of undetected pneumothorax cases could be delayed by already a small FPR.
With respect to the overall performance of the individual methods, the CNN stands out, whereas the FCN allows for the detection of $57\%$ of all findings with $1\%$ false alarms, only exceeded by the ensemble with a TPR of $68\%$.

While image-level annotations are most convenient for the development, for algorithms such as CNNs and MIL, the most relevant features for the discrimination of the different images are identified in an optimization process.
As a result, there is a substantial risk that non-relevant features, which are strongly correlated to the presence of a disease can contribute to the decision. 
In a recent study using the NIH ChestX-ray14 dataset, it was demonstrated that a CNN learned not only to detect the presence of a pneumothorax, but also of drains, which are frequently employed for treatment purposes \cite{baltruschat2018comparison}. 

On the other hand, the FCN approach requires pixel-level annotations. These are usually difficult to obtain, but the network provides a localization of the pneumothorax, which forms an additional level of confidence and interpretability.

Finally, both the CNN as well as the MIL approaches make use of the pre-trained network architectures, which require massive amounts of data for training. Availability of data in medical imaging applications is often limited, which makes the use of such pre-trained networks more appealing. Should such pre-trained networks be available for 3D applications, our approach could be extended for 3D applications, e.g.~pneumothorax detection in CT images.

\section{Conclusion}
The three presented techniques provide promising options for the detection and localization of pneumothorax in chest X-ray images.

We achieved the best performance in terms of AUC using CNN, whereas the MIL and FCN provided higher confidence in terms of localization. This could guide radiologists by visualizing the image region responsible for the network's decision, while simultaneously increasing the trust in the proposed deep learning architecture. 
Combining the proposed three methods as an ensemble, increased the overall classification performance, while MIL and FNC allow for a localization of the pathology. Future work could elaborate on other techniques to combine the three approaches, e.g.~by cascading networks or merging the architectures into one multi-task network.

\section*{Acknowledgments}
We would like to thank Christopher Hall for his support and advice. We further thank Tom Brosch and Rafael Wiemker for annotating data and providing valuable input on our network architectures. Finally, we thank Hannes Nickisch for reviewing the manuscript and providing valuable feedback.

\bibliographystyle{splncs04}
\bibliography{pneumo_paper}
\end{document}